\documentclass[12pt]{article}
 \usepackage{moreverb} 
\usepackage{cite}
\usepackage{graphicx}
\usepackage{subfig}
\usepackage{float}
\usepackage{subfloat}
\usepackage{epsfig}
\RequirePackage{amssymb, amsfonts, mathrsfs, amsmath, latexsym, verbatim, xspace, setspace}
\usepackage{longtable}
\usepackage{hyperref}
\usepackage{authblk}

 \hypersetup{colorlinks = true,
             linkcolor = red,
             anchorcolor = green,
             citecolor = blue,
             filecolor = red,
             pagecolor = red,
             urlcolor = red}

\title{A cosmological model  with time dependent $\Lambda$, $G$ and viscous fluid in General Relativity }
\author[1]{Rishi Kumar Tiwari \thanks{rishitiwari59@rediffmail.com}}
\author[2]{Alnadhief H. A. Alfedeel \thanks{aaalnadhief@imamu.edu.sa}}
\author[3]{De\u{g}er Sofuo\u{g}lu  \thanks{degers@istanbul.edu.tr}}
\author[4, 5]{Amare Abebe\thanks{AmareAbebe.Gidelew@nwu.ac.za}}
\author[2]{Eltegani I. Hassan \thanks{eiabdalla@imamu.edu.sa}}
\author[1]{B. K. Shukla \thanks{bhupendrashukla121@gmail.com}}
\affil[1]{Department of Mathematics, Govt. Model Science College Rewa (M.P.) India}
\affil[2]{Department of Mathematics and Statistics,  Imam Mohammad Ibn Saud Islamic University (IMSIU), Riyadh 13318, Saudi Arabia}
\affil[3]{Department of Physics, Istanbul University, 34134, Vezneciler, Fatih, Instanbul, Turkey} 
\affil[4]{Center for Space Research, North-West University, Mahikeng 2745, South Africa}
\affil[5]{National Institute for Theoretical and Computational Sciences (NITheCS), South Africa}

\begin{document}

\maketitle

\begin{abstract} 
  In this paper, we investigate Bianchi type$-V$ cosmological models with bulk viscous fluid and time varying cosmological $\Lambda$ and Newtonian $G$ parameters. The Einstein's field equations have been transformed into a coupling non-linear, 
 first-order differential equations, and the fourth-order Runge-Kutta  method of numerical integration has been used to integrate the differential equations with appropriate initial conditions consistent with current cosmological observations. We show that the model describes a universe that starts off with a negative cosmological term, as well as a matter-dominated and decelerated early epoch that, eventually becomes $\Lambda$-dominated and expanding with acceleration, in concordance with current observations.
\end{abstract}

\textbf{Keywords}: viscous fluid; bulk viscosity; Bianchi type-$V$; varying $\Lambda$ and $G$

\section{Introduction}
Bianchi cosmological models are homogeneous and anisotropic models that can be viewed
as a generalization of the homogeneous and isotropic Friedman-Lema\^itre-Robertson-Walker (FLRW) 
{\color{black} space-times} on which the concordance cosmology is based. These models are interesting because, although the universe is almost isotropic on the largest possible scales, small-scale anisotropies are a feature of the the observed universe. Current cosmological observations \cite{ perlmutter1997measurements, perlmutter1998discovery, perlmutter1999measurements, riess1998observational} point
out that the universe is {\color{black} expand with acceleration } that was previously thought to be decelerating.
Dark energy (DE), which in the standard $\Lambda$-Cold Dark Matter ($\Lambda$CDM) paradigm is represented by the cosmological constant in Einstein Field Equations (EFEs) of general relativity (GR), is thought
to be responsible for the late-time accelerated expansion. Whereas DE is estimated to account for about 70\% of the total matter-energy budget of the universe, the significant other proportion of 25\% is thought to exist in the form of dark matter (DM), a non-luminous, and yet-to-be discovered, form of matter whose presence can only be detected through its gravitational effects.

Several proposals have been put forward to address the issues of DE and DM, such as additions of exotic matter forms (such as $\Lambda$,  the Chaplygin gas, scalar fields, etc), or modifications of GR itself (in the form of $f(R)\;, f(T)\;, f(R,T)\;, f(Q)$, etc 
\cite{sotiriou2010f,nojiri2004modified,bertolami2007extra,bergliaffa2006constraining,bertolami2009energy,moraes2019cosmological,bamba2017energy,liu2012energy,atazadeh2014energy, Sharif2016,wang2012energy,mandal2020energy,arora2021constraining}) or the non-relativistic limit of Newtonian theory (MOND) in the case of DM. Another consideration in the literature has been one based on Dirac's hypothesis on the evolution of the fundamental `constants'.
P. Dirac hypothesized that $\Lambda$ must be a time-dependent function \cite{Dirac1937} because the
theoretical prediction from quantum field theory (QFT) differs significantly from observations in
the value of $\Lambda$ \cite{sahni2000case, chen1990implications}. Since then, various scientists have indicated an interest in investigating cosmological models in the context of GR with time-dependent cosmological constant $\Lambda$.
Many authors have investigated different forms of $\Lambda$ with standard and non-standard cosmological
models based on the same assumption \cite{Alfedeel:2018thd,alfedeel2020bianchi, Vishwakarma1996a, Vishwakarma1996b, Vishwakarma1999, Vishwakarma2000, Vishwakarma2001, Vishwakarma2005,bali2012bianchi}. 
Bulk viscosity is important in cosmology because it plays a role in the universe's accelerated expansion, also known as the inflationary phase. 
Over the course of the universe's history, bulk viscosity could manifest in a variety of ways \cite{ellis1971general}.
It is believed that viscosity emerges when neutrinos disengage from the cosmic fluid \cite{misner1968isotropy}, at the time of
galaxies formation and particle synthesis at the early stages of our cosmos \cite{hu1983advances}.
For all these reasons, there have been many attempts to study non-standard cosmological models involving viscous fluids.
Several researchers have recently studied various Bianchi-type cosmological models with varying cosmological constant ($\Lambda$) and bulk viscous fluid, including
\cite{tiwari2017cosmological, tiwari2017scenario,tiwari2016behaviour, 
tiwari2018cosmological,tiwari2018scenario, Arbab1997, Arbab1998,huang1990anisotropic,bali2007bianchi,bali2008bulk}.
For example, the influence of bulk viscosity on cosmic evolution has been studied in \cite{huang1990anisotropic,bali2007bianchi,bali2008bulk}.
Singh et al. \cite{singh2016viscous} investigated the Bianchi type$-V$ cosmological models for a viscous fluid, assuming that the Hubble parameter $H$ is a linear hyperbolic function of cosmic time $t$. They discovered that using the proposed functional form for the Hubble parameter results in cosmological models that are compatible with current observations.
Bali et al. \cite{bali2012bianchi} studied the Bianchi type$-V$ cosmological model for viscous fluid distribution with variable cosmological term $\Lambda$.
They analyzed a cosmic scenario after assuming the rule of variation for the Hubble parameter $H$, i.e. $H=a(R^{-n} + 1)$, where $a, n$ are constants and $R$ is the average scale factor. They discovered that the model isotropizes asymptotically, and that the existence of shear viscosity speeds up the isotropization. Singh and Baghel \cite{singh2010bulk} have investigated Bianchi type$-V$ cosmological models in the presence of bulk viscosity.
They derived an accurate solution for the EFEs by assuming that the shear scalar $\sigma$ is proportional to the volume expansion $\theta$, and that the coefficient of bulk viscosity is a power function of energy density $\rho$ or volume expansion $\theta$. They discovered that $\Lambda$ should be negative, and the models derived are expanding, shearing, and non-rotating, with no approach to isotropy at late periods.
The same authors analyzed spatially homogenous and anisotropic Bianchi type$-V$ space-times with a bulk viscous fluid source and a time-dependent cosmological term \cite{singh2009bianchi}. They arrived to cosmological models by assuming a law of variation for the Hubble parameter, which results in a constant deceleration parameter $q=m-1$, where $m$ is a constant. They came to the conclusion that the model reflected the universe's accelerating phase for particular values of $m$..
Padmanabhan and Chitre \cite{padmanabhan1987viscous} looked at the influence of bulk viscosity on the development of the cosmos as a whole. They demonstrate that the bulk viscosity can result in inflation-like solutions.

Motivated by the above discussion, in this paper we will investigate the Bianchi type$-V$ cosmological model for bulk viscous universe with time-dependent
cosmological parameter $\Lambda$ and Newtonian gravitational parameter $G$, which is inspired by previous works as mentioned above. We will not
assume any coupling relation between the metric variables in this study when solving the gravitational
field equations for model physical parameters or imposing any extra constraints, as others do.
Instead, we will recast the geovering equations for the Bianchi type$-V$ model as adimension less, non-linear, first
order, coupling differential equation for cosmological observations $h(z), \Omega_m(z),\Omega_\Lambda,\Omega_\chi$, and $\Omega_\sigma$, then
integrate them in parallel to estimate the other model characterized parameters.
The following is how the rest of this paper is organized: Section \ref{efes} introduces
the Bianchi type$-V$ metric and the field equations that go with it. The solution to the field equation
is presented in Section \ref{numerics}. Section \ref{resdis} will offer several cosmological models based on the selection of
time-varying shear and bulk viscosity. Finally, we bring the article to a close with our conclusion in Section \ref{conc}.

\section{Metric and field equations}\label{efes}
The Bianchi type$-V$ line-element in orthogonal space and time coordinates is represented
by the following formula:
\begin{equation}
ds^2=dt^2-A^{2}dx^2-e^{2m x}[B^{2}dy^2+C^{2}dz^2]~.\label{metric}
\end{equation}
where $A=A(t)$, $B=B(t)$ and $C=C(t)$ are the metric potential and $m$ is constant.
We assume that the universe is filled by a viscous fluid whose distribution
in space is represented by the following {\color{black} energy-momentum} tensor:
\begin{equation}
T_{ij} = (\rho  + \bar{p}) v_{i} v_{j} + \bar{p} g_{ij} - 2 \eta \sigma_{ij}~,\label{Tij} 
\end{equation}
where $\rho$ is matter energy density, $p$ is the isotropic pressure, $\eta$ and $\xi$ are coefficient
of shear and bulk viscosity respectively, $v^i = (v^1,v^2,v^3,v^4) = (0,0,0,1) $ is
$4$-velocity vector of the cosmic fluid and it is time-like
quantity that satisfying $ v_i v^i = -1$, $\sigma_{ij}$ is the shear
and $\bar{p}$ is the effective pressure which is given by
\begin{align}
\bar{p} = p - \xi v_{i;i}= p - \left(3 \xi - 2 \eta \right)H~.\label{preasure} 
\end{align}
{\color{black}Note that the bulk and shear viscosities, $\xi$ and $\eta$, are both positive, i.e., $\eta>0, \xi>0$. We will
assume them as either constant or function of time or energy, such as $\eta \sim H$ and $\xi \sim \rho^n$ and $n$ is a numerical constant.}
Here, the cosmic fluid is assumed to satisfy a linear equation of state
\[ p = w \rho~,\qquad -1 \leq w \leq 1~,\]
where $w$ is the equation of state parameter (EoS) which relates $p$ to
the energy density. The shear tensor is given by
\begin{align}
\sigma_{ij} = \left(v_{i;k} h_j^k + \dot{v}_{j;k} h_i^k \right) - \frac{1}{3} \theta h_{ij}~,\label{sheartensor}
\end{align}
where $h_{ij} = g_{ij} + v_i v_j$ is the projection tensor.
The Einstein field equations (EFEs) of the gravitation with time-varying cosmological
constant ($\Lambda$) in geometrical units where $ c = 1$ are given by
\begin{equation}
R_{ij} - \frac{1}{2}g _{ij} R =- \kappa G T_{ij}+ \Lambda g_{ij}\;.\label{EinsteinEq}
\end{equation}
Here  $\kappa\equiv 8\pi$ and $R_{ij}$ is Ricci tensor, $R$ is Ricci scalar and $g_{ij}$ is the symmetric
second-rank metric tensor. Using Eqs. \eqref{metric}-\eqref{sheartensor}, the 
EFEs in \eqref{EinsteinEq} for a viscous fluid distribution reduce to the following set of 
partial differential equations\footnote{Overdots represent partial differentiation with respect to cosmic time $t$.}:
\begin{align}
\frac{m^2}{A^{2}} -  \frac{\ddot{B}}{B} - \frac{\ddot{C}}{C}
- \frac{\dot{B}}{B}\frac{\dot{C}}{C} + 2 \eta \frac{\dot{A}}{A} & = 
\kappa G \left[ p - \left( \xi - \frac{2}{3} \eta \right) \theta \right] -\Lambda~,\label{Gtt}\\
\frac{m^2}{A^{2}} -  \frac{\ddot{A}}{A} -\frac{\ddot{C}}{C} 
- \frac{\dot{A}}{A}\frac{\dot{C}}{C} + 2 \eta \frac{\dot{B}}{B} & = 
\kappa G \left[ p - \left( \xi - \frac{2}{3} \eta \right) \theta \right] -\Lambda~,\label{Gxx}\\
\frac{m^2}{A^{2}} -  \frac{\ddot{A}}{A} -\frac{\ddot{B}}{B} 
- \frac{\dot{A}}{A}\frac{\dot{B}}{B} + 2 \eta \frac{\dot{C}}{C} & = 
\kappa G \left[ p - \left( \xi - \frac{2}{3} \eta \right) \theta \right]  -\Lambda~,\label{Gyy}\\
- \frac{3m ^{2}}{A^{2}} +  \frac{\dot{A}}{A}\frac{\dot{B}}{B}
+\frac{\dot{A}}{A}\frac{\dot{C}}{C} +\frac{\dot{B}}{B}\frac{\dot{C}}{C}
& = \kappa G \rho +\Lambda~,\label{Gzz}\\
\frac{\dot{B}}{B}+\frac{\dot{C}}{C}-2\frac{\dot{A}}{A} & =0~.\label{Gxz}
\end{align}

{\color{black} Generally, one can consider that the covariant derivative of the energy-momentum tensor $T_{ij}$ is proportional to the time variation of the cosmological `constant' and the gravitational `constant', thus:}
\begin{align}
\kappa G\left[ \dot{\rho} + (\overline{p} + \rho)\left( \frac{\dot{A}}{A} + \frac{\dot{B}}{B}  
+ \frac{\dot{C}}{C}  \right)\right ] + \kappa \rho \dot{G} + \dot{\Lambda}
- 4\kappa G \eta \sigma^2=0\;.\label{ConsEq} 
\end{align}
{\color{black}
Using $\overline{p} =p -(3\xi - 2\eta) H$, 
if the total matter content of the universe is conserved,
Eq. \eqref{ConsEq} can be split into two independent equations:}
\begin{eqnarray}
&&\dot{\rho} + 3 H \left[p + \rho - (3\xi - 2\eta) H\right]- 4\eta \sigma^2 =  0~, \label{Rho+Evol}\\
&&\kappa \rho \dot{G}  + \dot{\Lambda} = 0  \;.\label{GL+Evol}
\end{eqnarray}
{\color{black}
According to Eq. \eqref{GL+Evol}, $G$ turns out to be constant for non-zero energy density $\rho$ when $\Lambda$ is constant or $\Lambda = 0$.} 
Note that we have used $H \equiv 1/3\left( \frac{\dot{A}}{A} + \frac{\dot{B}}{B}
+ \frac{\dot{C}}{C}  \right)$ as we will show later, and $\sigma$ is the scalar shear tensor
 is given by
\begin{align}
\sigma^2 = \frac{1}{2}\sigma_{ij}\sigma^{ij}=\frac{\sigma_0^2}{a^6}\;,\label{shearEq} 
\end{align}
where $\sigma_0$ is a constant that is related to the universe anisotropy. The spatial volume $V$ for Bianchi type$-V$ space-time given by 
\begin{align}
V = a^3 = \sqrt{|-g_{ij}|} = ABC~,
\label{Volume}
\end{align}
where $(a)$ is the average scale factor of universe. In addition to
that, the generalized Hubble parameter $H$, and the
deceleration parameter $q$ are defined as
\begin{equation}
H\equiv \frac{\dot{a}}{a}=\frac{1}{3}\left(H_{x}+H_{y}+H_{z}\right)\;,\quad q\equiv -\frac{a\ddot{a}}{\dot{a}^{2}}=-\frac{\dot{H}}{H^{2}}-1 ~,\label{qdef} 
\end{equation}
where $ H_{x}\;, H_{y}$ and $H_{z}$ are the directional Hubble parameters
along $x,y$ and $z$ directions respectively. The components
of the shear tensor $\sigma_{ij}$
for the metric in Eq. \eqref{metric} are calculated as 
\begin{equation}
\sigma_{11}  =H_{x}-H\;,\quad\sigma_{22} =H_{y}-H\;,\quad  \sigma_{33} =H_{z}-H\;,\quad \sigma_{44}=0~,\label{sigma+components}
\end{equation}
and the shear scalar $\sigma$ now gives  
\begin{equation}
\sigma^{2}=\frac{1}{6}\left[ \left( \frac{\dot{A}}{A}
-\frac{\dot{B}}{B}\right) ^{2}
+\left( \frac{\dot{B}}{B}-\frac{\dot{C}}{C}\right) ^{2}
+\left( \frac{\dot{C}}{C}-\frac{\dot{A}}{A}\right)^{2}\right]~.\label{shear+square}
\end{equation}
The average anisotropy parameter $A_p$ is defined as
\begin{equation}
A_p   = \frac{1}{3}\sum_{i=1}^{3} \left(\frac{H_i - H}{H} \right)^2\;.
\end{equation}
Subtracting the field equations \eqref{Gxx} and \eqref{Gyy} gives
\begin{equation}
\frac{\ddot{B}}{B}-\frac{\ddot{C}}{C}+\left( \frac{\dot{B}}{B}-\frac{\dot{C}}{C}\right) \left\lbrace 
\frac{1}{2} \left( \frac{\dot{B}}{B}+\frac{\dot{C}}{C}\right) +2\eta \right\rbrace  =0\;,
\end{equation}
which can be integrated to give 
\begin{equation}
\frac{\dot{B}}{B}-\frac{\dot{C}}{C} = \frac{k_1}{a^{3}}e^{-2\int \eta dt}~,\label{BC}
\end{equation}
Similarly, the rest of the field equations \eqref{Gtt}-\eqref{Gxz}
can also be solved to give a coupled first order differential equation
for the metric variables $A$, $B$ and $C$ as
\begin{align}
\frac{\dot{A}}{A} & =\frac{\dot{a}}{a}~,\label{AaEq} \\
\frac{\dot{C}}{C}-\frac{\dot{A}}{A}& = \frac{k_2}{a^{3}}e^{-2\int \eta dt}\;.\label{CA}
\end{align}
Integrating Eq. \eqref{AaEq} and
absorbing the constant of integration into $A$ or $B$ yields
\begin{align}
 A=a\;. \label{AEq}
\end{align}
Thus, plugging Eq. \eqref{AEq} into Eq. \eqref{BC} and \eqref{CA} produces
\begin{align}
\frac{\dot{B}}{B} & =\frac{\dot{a}}{a} + \frac{k_1}{ a^{3}} e^{-2\int \eta dt}~,\label{BcEq}\\
\frac{\dot{C}}{C} & =\frac{\dot{a}}{a} + \frac{k_2}{a^{3}}e^{-2\int \eta dt}\;.\label{Caq}
\end{align}
Integrating these equations one more time gives an expression for the metric function $B$ and $C$ as 
\begin{align}
B  & = d_1 a \exp\left[ \int \left( \frac{k_1}{ a^{3}} e^{-2\int \eta dt} \right) dt  \right] ~,\label{BEq}\\
C  & = d_2 a \exp\left[ \int \left( \frac{k_2}{ a^{3}} e^{-2\int \eta dt} \right) dt  \right]~,\label{CEq}
\end{align}
where $ k_1$, $k_2$, $d_1$ and $d_2$ are constants of integration.  
Equations \eqref{Gtt}-\eqref{Gxz} can be written in terms of $H$, $\sigma$ and $ q $ as
\begin{eqnarray}
&&\kappa G \overline{p}-\Lambda = H^{2}(2q-1)-\sigma^{2} +\frac{m^{2}}{A^{2}}~,\label{Friedman1}\\
&&\kappa G \rho +\Lambda =3H^{2}-\sigma^{2} -\frac{3m^{2}}{A^{2}}\;.\label{Friedman2}
\end{eqnarray}
Eq. \eqref{Friedman1} and Eq. \eqref{Friedman2} are the generalized
Friedmann equations for Bianchi type-$V$ spacetimes endowed with the viscous-fluid model under consideration. The generalized Raychaudhuri equation reads:
 \begin{align}
\dot{H} + 3H^2 - \frac{2m^2}{a^2} - \Lambda + \frac{\kappa G }{2}(p - \rho) 
- \kappa G \left( \frac{3\xi}{2} - \eta\right)H=0\;.\label{dHdtEq}
\end{align}
This equation cannot be solved as it stands because of the unknown variables $\eta, \xi,a,G,\Lambda,p$ and $\rho$. In order to facilitate the solution process by providing extra information in the form of initial conditions and a constraint,  we divide the re-arranged form of the Friedmann equation \eqref{Friedman2} by $3H^2$ and write
\begin{equation}\label{constr}
1 =  \Omega_m + \Omega_\Lambda +  \Omega_\sigma + \Omega_\chi
\end{equation}
such that 
\begin{align}
 \Omega_{m}\equiv \frac{\kappa G \rho_{m}}{3H^2}~, \qquad
\Omega_{ \Lambda} \equiv \frac{\kappa G \rho_{\Lambda }} {3H^2}~,\qquad 
\Omega_{\sigma}\equiv \frac{\sigma^2}{ 3H^2}~, \qquad 
\Omega_{\chi}\equiv \frac{ 3 m^2 }{ 3H^2 a^2 }\;.
\end{align}
The present-day values of the above dimensionless quantities are given by
\begin{align}
 \Omega_{m_0}= \frac{\kappa G_0 \rho_{m_0}}{3H^2_0}~, \qquad
\Omega_{ \Lambda_0} = \frac{\kappa G_0 \rho_{\Lambda_{0} }} {3H^2_0}~,\qquad 
\Omega_{\sigma_0}= \frac{\sigma^2_{0} }{ 3H^2_0}~, \qquad 
\Omega_{\chi_0}= \frac{ 3 m^2 }{ 3H^2_0 a_0^2 }~,
\end{align}
In terms of the dimensionless parameters defined here,  Eqs. \eqref{Rho+Evol} and \eqref{GL+Evol} can be rewritten as: 
\begin{align}\label{omdot}
\dot{\Omega}_m & + \left( 2\frac{\dot{H}}{H} -\frac{\dot{G}}{G} \right)\Omega_m 
               + 3H\left[ (1+w_m)\Omega_m - \frac{\kappa G}{3H} (3\xi -2\eta) \right] -4 \kappa G \eta \Omega_\sigma =0\;,\\              
\dot{\Omega}_\Lambda & +  2\frac{\dot{H}}{H} \Omega_\Lambda +\frac{\dot{G}}{G}  	\Omega_m   =0\;.\label{oldot}
\end{align}
These evolution equations together with the constraint \eqref{constr} need one extra equation to solve for the different $\Omega_i$'s. Thus we give the following additional evolution equations for the fractional energy density of the:
\begin{eqnarray}\label{curevol}
&&\dot{\Omega}_\chi+2\left(H+\frac{\dot{H}}{H}\right)\Omega_\chi=0\;,\\
&&\dot{\Omega}_\sigma+\left(6H+2\frac{\dot{H}}{H}\right)\Omega_\sigma=0\;.
\end{eqnarray}
Our next step is to numerically integrate these equations and see if/how the results compare
with those of the $\Lambda CDM$ model.

\section{Numerical Integration}\label{numerics}
We observe that a viscous fluid Bianchi type-$V$ model with time varying $G$ and $\Lambda$ is characterized
by $A,B,C$, $h$. $q$, $\Omega_m$, $\Omega_{\Lambda}$ and $G$, but the system of equations
Eqs. \eqref{Gtt}-\eqref{Gxz}, \eqref{Rho+Evol} and \eqref{GL+Evol} only provides five differential
equations. To complete the solutions processes an extra equation or assumption is required.
According to the Dirac \cite{Dirac1937} ansatz, the gravitational constant must
decrease with time, and based on this we assume that    
\begin{align}
G(t) = {\color{black} G_0 a^{\delta} ~\implies \dot{G} = G_0 \delta H}~,\label{GAssumption}
\end{align}
{\color{black} where $\delta = -1/60$ is a constant obtained from observational constraints \cite{williams2009lunar}}.
In order to transform the governing Bianchi type-$V$ evolution equations in redshift space, we use 
\begin{align}
\dot{Q}= \frac{dQ}{dt} &= \frac{dQ}{dz} \; \frac{dz}{da}\; \frac{da}{dt} = -(1+z)HQ'
 \end{align}
for any time-dependent quantity $Q$,  and 
with the dimensionless parameters
\[ h\equiv \frac{H}{H_0}\;, \qquad a =\frac{1}{(1+z)}~,\qquad \xi=  \alpha H_0(\rho_{\rm m}/\rho_{\rm m0})^n~, \qquad \mbox{and} ~~~\eta = \beta H\;. \]
Here $\alpha$ and $\beta$ are dimensionless constants and $0 \leq n \leq \frac{1}{2}$.  We can thus rewrite our previous equations \eqref{dHdtEq}, \eqref{omdot},\eqref{oldot} and \eqref{curevol}
 in fully dimensionless forms as follows:
\begin{align}
h' &= \frac{h}{(1+z)} \left[ 3 - 2 \Omega_\chi - 3\Omega_\Lambda -  \frac{3}{2} (1-w_m) \Omega_m  
               \right] { -\left[  \frac{3 \alpha \kappa G_0 }{2} \left( \frac{ h^2 \Omega_m}{\Omega_{m0}} \right)^n - \kappa G_0 \beta h  \right] \frac{1}{(1+z)^{1+\delta}} }  \label{dhdzDE} \\               
 \Omega'_m  & =   -\frac{2h'}{h} \Omega_m  + \frac{1}{1+z}  \left(4
               + 3w_m\right)\Omega_m
               - \frac{3 \alpha \kappa G_0 }{h {\color{black} (1+z)^{1+\delta} }  } \left( \frac{h^2 \Omega_m}{ \Omega_{m0}}\right)^n + \frac{2 \kappa G_0 \beta}{{\color{black} (1+z)^{1+\delta}}} -  \frac{4 \beta \kappa G_0 \Omega_\sigma}{{\color{black} (1+z)^{1+\delta}}}\;,
               \\
 \Omega'_\Lambda& =   -\frac{2h'}{h} \Omega_\Lambda  - \frac{{\color{black} \delta}}{1+z}  \Omega_m
                \;,\label{dLdzDE}\\          
                \Omega'_\chi& = -\frac{2h'}{h} \Omega_\chi  + \frac{2\Omega_\chi}{1+z}\;,\label{curvevz}\\
                \Omega'_\sigma&=-\frac{2h'}{h} \Omega_\sigma  + \frac{6\Omega_\sigma}{1+z}\;,\label{sigevz}
                \end{align}
 Eqs. \eqref{dhdzDE}-\eqref{sigevz} are first-order
coupled differential equations that describe the evolution of $h$, $\Omega_m$ and $\Omega_\Lambda$
with respect to the redshift $z$. The deceleration parameter $q$, the metric variables
$A$, $B$ and the volume expansion $V$ are given by:
\begin{align}
q&= 2 - 2 \Omega_\chi - 3\Omega_\Lambda -  \frac{3}{2} (1-w_m) \Omega_m  
    - \frac{3 \alpha \kappa G_0 }{2h (1+z)^{\color{black} \delta}} \left( \frac{ h^2 \Omega_m}{\Omega_{m0}}\right)^n + \frac{\beta \kappa G_0}{(1+z)^{\color{black} \delta}}\;,\\
B &= \frac{d_1}{(1+z)}\; \exp \left\{ \frac{\kappa_1}{H_0}  \int \frac{(1+z)^{2+2\alpha}}{h}  dz \right\}~,\label{BFormula}\\
C &= \frac{d_2}{(1+z)}\; \exp\left\{\frac{\kappa_2}{H_0}  \int \frac{(1+z)^{2+2\alpha}}{h} dz \right\}\;,\label{CFormula} \\
V &= ABC = \frac{d_3}{(1+z)^3}\; \exp\left\{\frac{\kappa_3}{H_0}  \int \frac{(1+z)^{2+2\alpha}}{h} dz \right\}\;,\label{VFormula}
\end{align}
where $d_3= d_1d_2$ and $\kappa_3= \kappa_1 + \kappa_2$ are numerical constants.

\begin{figure}[htp]
  \centering
  \subfloat[$\Omega_\Lambda$ vs z.]{\includegraphics[width=0.55\textwidth]{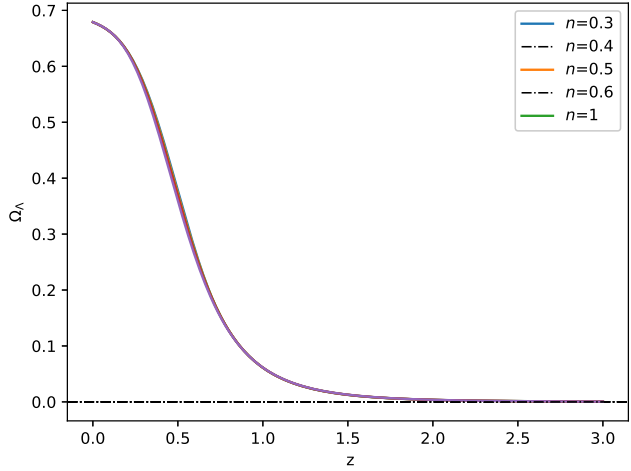}}
  \subfloat[$\Omega_{\rm m}$ vs z.]{\includegraphics[width=0.55\textwidth]{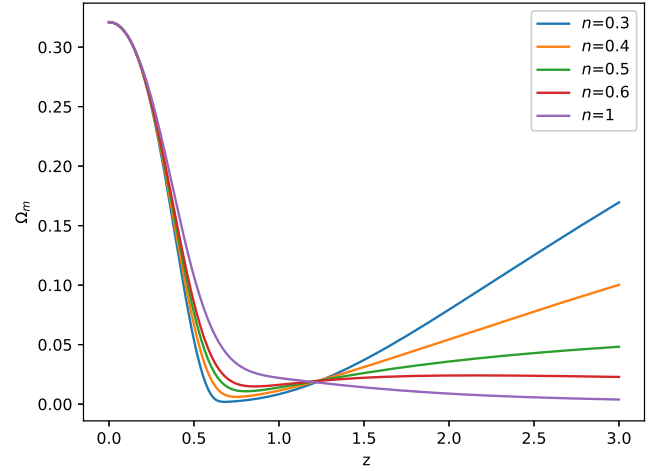}}
  \caption{The variation of the fractional energy densities of dark energy $\Omega_\Lambda$ and matter $\Omega_m$ with redshift.
  The current values from \cite{aghanim2020planck} $h(0) = 1, \Omega_{m} (0) \equiv \Omega_{\rm m0}= 0.321\;, \Omega_{\Lambda}(0          1     )\equiv \Omega_{\Lambda 0}=0.679\;, \Omega_{\rm \chi 0}=-0.056$ and $\Omega_{\rm\sigma 0}=1- \Omega_{\rm m0}- \Omega_{\Lambda 0}-\Omega_{\rm \chi 0}$ are used as initial conditions along with the fourth-order
Runge-Kutta method to integrate the system numerically.}
      \label{fig1}
\end{figure}
\begin{figure}[htp]          
       \centering
       \subfloat[$h$ vs z.]{\includegraphics[width=0.55\textwidth]{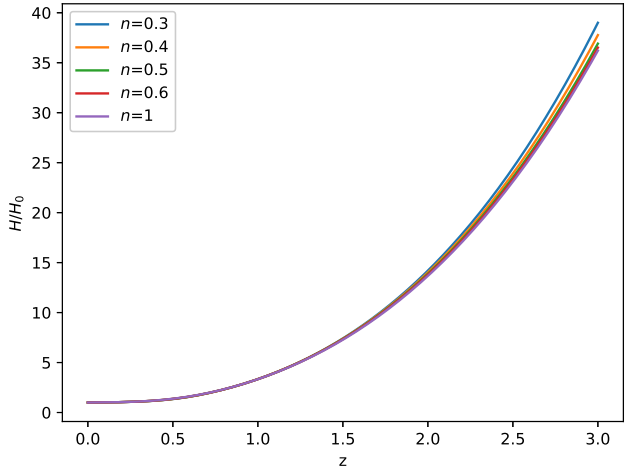}}
  \subfloat[$q$ vs z.]{\includegraphics[width=0.55\textwidth]{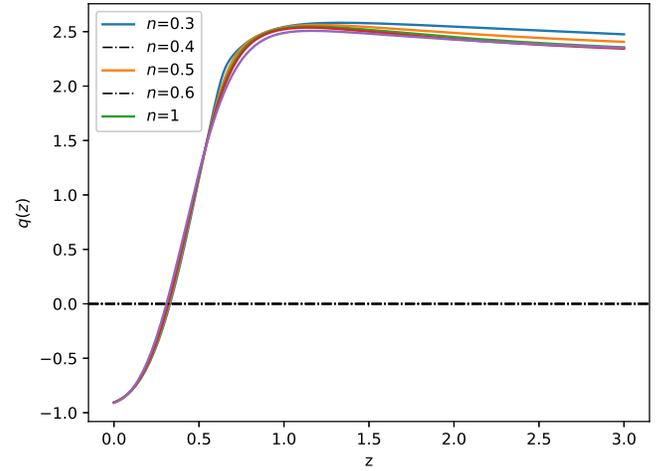}}
  \caption{The variations of the normalized expansion rate $h$ and the deceleration parameter $q$ with redshift.
  The current values from \cite{aghanim2020planck} $h(0) = 1, \Omega_{m} (0) \equiv \Omega_{\rm m0}= 0.321\;, \Omega_{\Lambda}(0)\equiv \Omega_{\Lambda 0}=0.679\;, \Omega_{\rm \chi 0}=-0.056$ and $\Omega_{\rm\sigma 0}=1- \Omega_{\rm m0}- \Omega_{\Lambda 0}-\Omega_{\rm \chi 0}$ are used as initial conditions along with the fourth-order
Runge-Kutta method to integrate the system numerically.}
\label{fig2}
\end{figure}

\begin{figure}[htp]
  \centering
  \subfloat[$\xi$ vs z.]{\includegraphics[width=0.55\textwidth]{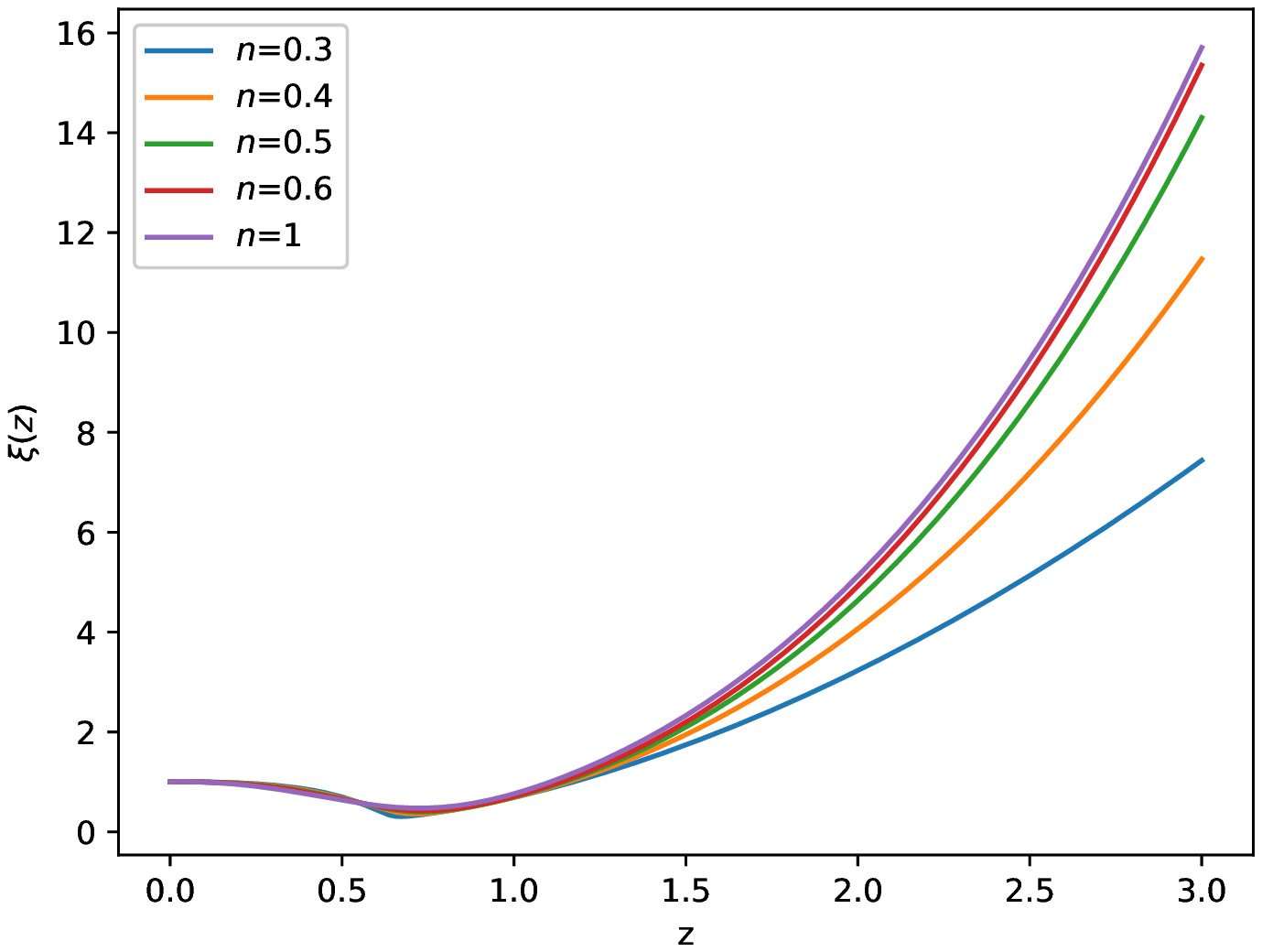}}
  \subfloat[$A_p$ vs z.]{\includegraphics[width=0.55\textwidth]{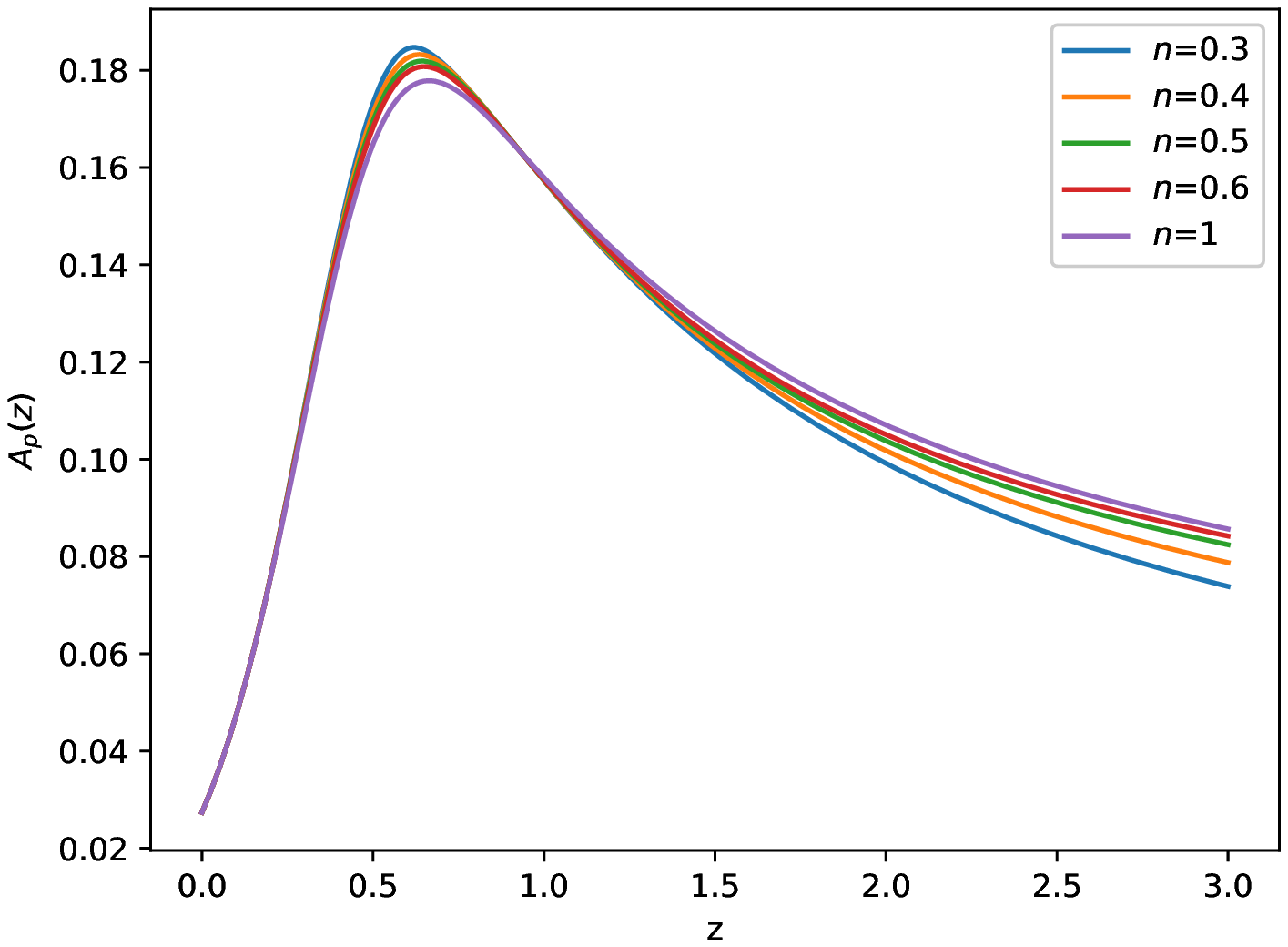}}
  \caption{The variation of the bulk viscosity$\xi$ and anisotropy $A_p$ parameters with redshift.
  The current values from \cite{aghanim2020planck} $h(0) = 1, \Omega_{m} (0) \equiv \Omega_{\rm m0}= 0.321\;, \Omega_{\Lambda}(0)\equiv \Omega_{\Lambda 0}=0.679\;, \Omega_{\rm \chi 0}=-0.056$ and $\Omega_{\rm\sigma 0}=1- \Omega_{\rm m0}- \Omega_{\Lambda 0}-\Omega_{\rm \chi 0}$ are used as initial conditions along with the fourth-order
Runge-Kutta method to integrate the system numerically.}
      \label{fig3}
\end{figure}

\section{Results and Discussion}\label{resdis}
The model governing system of equations  \eqref{dhdzDE}-\eqref{dLdzDE} is numerically solved for 
$h(z),\Omega_{m}$ and $\Omega_\Lambda$ along with the normalized initial conditions
$h(0) = 1, \Omega_{m} (0) \equiv \Omega_{\rm m0}= 0.321\;, \Omega_{\Lambda}(0)\equiv \Omega_{\Lambda 0}=0.679\;, \Omega_{\rm \chi 0}=-0.056$ and $\Omega_{\rm\sigma 0}=1- \Omega_{\rm m0}- \Omega_{\Lambda 0}-\Omega_{\rm \chi 0}$ using the fourth-order
Runge-Kutta method. The numerical results were obtained for several values of the
constant $n$ in the range $ 0\leq n \leq1$ and $\alpha=\beta = \kappa G_0=1$. The behaviors
of $\Omega_m\;, \Omega_\Lambda\;, h\;, q\;,\xi\,~\mbox{and}~A_p $ are graphically represented in Figs. \eqref{fig1}-\eqref{fig3}.

From Fig. \eqref{fig1} we see that $\Omega_m$ starts evolving with redshift from having large value at an earlier
stage of cosmic evolution gradually decreasing to its minimum value around $z\sim 1$, then reaching its current
value of $\Omega_{m0}$ at $z=0$, whereas $\Omega_\Lambda$ grew from a small value at
the early times to its current positive value at $z=0$. This result is in agreement with results
from the $\Lambda CDM$ model.

As seen in Figs. \eqref{fig2} and \eqref{fig3}, the normalized Hubble parameter $h$, the bulk viscosity $\xi$ and the anisotropy parameter $A_p$
have become smaller today compared to their values at larger redshifts, for all vales of n considered. It appears from our analysis, however, that the anisotropy term at about $z\sim 1$ (when the fractional energy density was at its minimum) reaches a maximum value before it decreases to its minimum value today.

The right panel of Fig.
\eqref{fig2} shows demonstrates that the deceleration parameter changes sign at small redshift values, from negative $q>0$ at
the early times to $q<0$ at the present time for all different values of $n$ considered. The change in $q$ indicates that the universe
expansion in this model has gone through a phase transition from slowing (decelerating) early epoch on to
a speeding up (accelerating) universe now, with the transition from deceleration to acceleration happening at $z\sim 0.5$, as predicted by observations as well.

\section{Conclusion}\label{conc}

The major goal of this paper was to investigate the homogeneous and anisotropic
Bianchi type$-V$ cosmological model in the presence of shear $\eta$ and bulk $\xi$
viscosities in the cosmic fluids for time-varying gravitational $G$ and cosmological $\Lambda$ parameters.
The governing background EFEs were simplified to second-order differential
equations for the metric variables $A, B, \& C$, as well as generalized Friedman equations.
In this research we have transformed the  basic governing equations into  non-linear
first-order differential equations for $h, \Omega_{\rm m}\;, \Omega_\Lambda\;, \Omega_\sigma\;, \Omega_\chi$ in the redshift space,  which may solved by numerically integrating in parallel using the fourth-order Runge-Kutta
method. Unlike previous studies that required a relationship between the model's characteristic
parameter to describe the model in time domain, the current method of integration is significant because
it allows us to determine the behaviour of the model directly from redshift-dependent measurable quantities and to compare it to current and future data. Our results showed that the model describes a universe that starts off with a negative cosmological term, dominated by non-relativistic matter and decelerated, that eventually becomes dark energy-dominated and hence expanding with acceleration, in concordance with current observations.
Our future endeavour in this direction will involve a more rigorous data analysis to observationally constrain the different assumed parameters of the model.

\section*{Acknowledgments}
The authors extend their appreciation to the Deanship of Scientific Research at Imam Mohammad
Ibn Saud Islamic University for funding this work through Research Group no. RG-21-09-18


\end{document}